\documentclass[12pt]{article}
\usepackage{geometry}                
\usepackage{graphicx}
\usepackage{amssymb}
\usepackage{amsmath}
\usepackage{epstopdf}
\usepackage{verbatim}

\def\ba{\begin{array}}
\def\ea{\end{array}}

\def\dalemb#1#2{{\vbox{\hrule height .#2pt
       \hbox{\vrule width.#2pt height#1pt \kern#1pt
               \vrule width.#2pt}
       \hrule height.#2pt}}}

\def\R{{{\Bbb R}}}

\def\Op{{\mathcal{O}}}

\def\rh{{\tilde{r}}}
\def\th{{\tilde{t}}}
\def\phih{{\tilde{\phi}}}

\newcommand{\be}{\begin{equation}}
\newcommand{\ee}{\end{equation}}
\newcommand{\bal}{\begin{align}}
 \newcommand{\eal}{\end{align}}
\newcommand{\ben}{\begin{equation*}}
\newcommand{\een}{\end{equation*}}
\newcommand{\bea}{\begin{eqnarray}}
\newcommand{\eea}{\end{eqnarray}}
\newcommand{\bean}{\begin{eqnarray*}}
\newcommand{\eean}{\end{eqnarray*}}
\newcommand{\bes}{\begin{subequations}}
\newcommand{\ees}{\end{subequations}}

\newcommand{\I}{{\mathcal R}}
\newcommand{\F}{{\mathcal F}}

\begin{document}

\begin{titlepage}
\bigskip
\rightline{}

\bigskip\bigskip\bigskip\bigskip
\centerline {\Large  {On the Stress Tensor of Kerr/CFT}}

\bigskip\bigskip
\bigskip\bigskip

\centerline{\large  Aaron J. Amsel,  Donald Marolf, and Matthew M. Roberts}
\bigskip\bigskip
\centerline{\em Department of Physics, UCSB, Santa Barbara, CA 93106}
\bigskip
\centerline{\tt  amsel@physics.ucsb.edu, marolf@physics.ucsb.edu, matt@physics.ucsb.edu}
\bigskip\bigskip

\begin{abstract}
The recently-conjectured Kerr/CFT correspondence posits a field theory dual to dynamics in the near-horizon region of an extreme Kerr black hole with certain boundary conditions.  We construct a boundary stress tensor for this theory via covariant phase space techniques.  The structure of the stress tensor indicates that any dual theory is a discrete light cone quantum theory, in agreement with recent arguments by Balasubramanian et al.  The key technical step in our construction is the addition of an appropriate counter-term to the symplectic structure, which is necessary to make the theory fully covariant and to resolve a subtle problem involving the integrability of charges.
\end{abstract}
\end{titlepage}

\tableofcontents

\setcounter{equation}{0}

\section{Introduction}

Holographic dualities are fascinating equivalences between quantum theories with gravity on the one hand and non-gravitational quantum field theories on the other.  The best established duality of this type is the anti-de Sitter/Conformal Field Theory (AdS/CFT) correspondence of string theory \cite{Maldacena:1997re}, which is connected to the near-horizon physics of certain supersymmetric asymptotically flat black holes.  It would be very interesting to understand similar correspondences involving  non-supersymmetric black holes.  A natural starting point is to consider the near-horizon geometry of an extreme Kerr black hole \cite{Zaslavsky:1997uu,Bardeen:1999px}, which was shown
in \cite{Bardeen:1999px} to be
described by a spacetime with isometry group  $SL(2, \mathbb{R}) \times U(1)$.   This spacetime is known either as the extremal (or extreme) Kerr throat \cite{Bardeen:1999px} or as the Near-Horizon Extreme Kerr (NHEK) geometry \cite{Guica:2008mu}.  We shall use these terms interchangeably below.

The asymptotic symmetry group of this spacetime was recently investigated in \cite{Guica:2008mu}, under a certain set of boundary conditions that we shall call GHSS fall-off.  The intriguing result was that this group contains a Virasoro algebra whose central charge is related via Cardy's formula to the entropy of the Kerr black hole.  As a result,  \cite{Guica:2008mu} conjectured that the physics of the near-horizon geometry of an extreme Kerr black hole  is dual to a 1+1 chiral CFT.  Similar results have now been shown to hold for very general extreme rotating black holes (see e.g. \cite{Azeyanagi:2009wf} and references therein for $d \ge 4$ as well as \cite{Balasubramanian:2009bg} for BTZ black holes).    One also finds \cite{Bredberg:2009pv} that scattering off of extreme Kerr black holes is consistent with a CFT description.

Roughly speaking, the CFT should be associated with the $t, \phi$ directions which describe the time-translation and rotation-symmetries of the extreme Kerr throat.  One therefore expects the objects which describe the CFT (for example, the stress tensor) to be covariant with respect to diffeomorphisms of the form $t,\phi \rightarrow x^i(t, \phi)$, where $x^i$ are new boundary coordinates.  Yet, the original constructions of \cite{Guica:2008mu} are not of this form and give special status to $t= constant$ surfaces.  Indeed, as we will see, the symplectic norm defined in
\cite{Guica:2008mu} is not conserved when evaluated on more general surfaces.  We also identify a related problem involving integrability of the charges defined in \cite{Guica:2008mu}.

We show below that  both issues can be resolved by adding a suitable boundary counter-term to the symplectic structure.  Similar boundary counter-terms were previously studied in \cite{Compere:2008us,Amsel:2009rr}, where they were shown to be closely related to counter-terms in the action.   As a result, one may consider our work below as a first step toward holographic renormalization of the action for Kerr/CFT along the lines of e.g. \cite{Henningson:1998gx,Balasubramanian:1999re, deHaro:2000xn,Bianchi:2001de, Bianchi:2001kw,Skenderis:2002wp}.

Our counter-term is constructed so as to give no contribution to symplectic products evaluated on constant $t$ surfaces.  As a result,  it does not affect the primary conclusions reported in \cite{Guica:2008mu}; the main effect is to render them covariant, so that any surfaces may be used for the computations.  In addition, our counter-term also fixes the above-mentioned problem with integrability.  The covariant description allows us to compute a covariant differential stress tensor describing the difference between two nearby solutions. This differential stress tensor is essentially a boundary analogue of the so-called canonical stress tensor familiar from scalar field theories (see e.g. \cite{LL}).  Our analysis is greatly simplified by making use of the conjecture of \cite{NoDynamics,Dias:2009ex} that all solutions satisfying GHSS fall-off are diffeomorphic to the extreme Kerr throat.

We begin with a brief review of the NHEK geometry in section \ref{review}.  We then analyze the phase space in section \ref{cps}, finding the required counter-term for the symplectic structure.  With a truly covariant phase space in hand, we then construct the boundary stress tensor in section \ref{bst}.  We close with some interpretation and a brief discussion of other issues in Kerr/CFT.

\setcounter{equation}{0}

\section{The extremal Kerr throat}
\label{review}

To orient the reader and establish conventions,
we begin by recalling how the extremal Kerr throat can be obtained as a scaling limit of the Kerr geometry \cite{Bardeen:1999px}.  The general Kerr metric is labeled by two parameters, a mass $M$ and angular momentum $J=Ma.$  The resulting black hole has temperature
\begin{equation}
\tilde T=\frac{\sqrt{M^2-a^2}}{4\pi M(M+\sqrt{M^2-a^2})}
\end{equation}
 and entropy $S=2\pi M(M+\sqrt{M^2-a^2})$.  In Boyer-Lindquist coordinates $(\th,\rh,\theta,\phih)$, the metric takes the form
\be
ds^2=-e^{2\nu}d\th^2+e^{2\psi}(d\phih + A d\th)^2+\Sigma(d\rh^2/\Delta+d\theta^2) \,,
\ee
where
\be
\Sigma=\rh^2+a^2\cos^2\theta,~\Delta=\rh^2-2M\rh+a^2 \,,
\ee
\be
e^{2\nu}=\frac{\Delta\Sigma}{(\rh^2+a^2)^2-\Delta a^2\sin^2\theta},~
e^{2\psi}=\Delta\sin^2\theta e^{-2\nu},~
A= -\frac{2M\rh a}{\Delta\Sigma}e^{2\nu}.
\ee
For the extremal solution, $a=M$ and $S=2\pi M^2=2\pi J$.   Note that we have set Newton's constant $G = 1$.

Defining a one-parameter family of new coordinate systems
\be
\rh=M+\lambda r,~\th=t/\lambda,~\phih=\phi+t/2M\lambda \label{nhekscaling}
\ee
and taking the scaling limit $\lambda\rightarrow 0$ yields
\be
ds^2=\left(\frac{1+\cos^2\theta}{2} \right)\left[ -f dt^2+dr^2/f+r_0^2d\theta^2\right]+\frac{2r_0^2\sin^2\theta}{1+\cos^2\theta}(d\phi+ r/r_0^2 dt)^2\, ,\label{nhekmetric}
\ee
with $r_0^2=2M^2$ and $f=r^2/r_0^2$.  This spacetime is known either as the extremal Kerr throat or as the Near-Horizon Extreme Kerr (NHEK) geometry.    For fixed $\theta$, the term in square brackets becomes the metric on $AdS_2$ in Poincar\'e coordinates.  As a result, we refer to  $(t, r, \theta, \phi)$ as Poincar\'e coordinates for the extremal Kerr throat.

The throat geometry inherits many properties from the above-mentioned AdS${}_2$.  For example, a geodesically complete spacetime can be obtained by applying the coordinate transformation
\be
\label{PoincareToGlobal}
r= (1+r'{}^2)^{1/2}\cos t'+r',~t= \frac{(1+r'{}^2)^{1/2}\sin t'}{r},~\phi=\phi'+\log\left|\frac{\cos t'+r' \sin t'}{1+(1+r'{}^2)^{1/2}\sin t'} \right| ,
\ee
which takes Poincar\'e AdS${}_2$ to the standard global coordinates on AdS${}_2$ \cite{Bardeen:1999px}.  The result is just
\be
ds^2=\left(\frac{1+\cos^2\theta}{2} \right)\left[ -\left(1 + \frac{r^2}{r_0^2}\right) dt^2+\frac{dr^2}{1 + r^2/r_0^2}+r_0^2d\theta^2\right]+\frac{2r_0^2\sin^2\theta}{1+\cos^2\theta}(d\phi+ r/r_0^2 d t)^2\, ,\label{nhekglobal}
\ee
where we have dropped the primes on coordinates in (\ref{nhekglobal}) for simplicity.  The result is precisely of the form (\ref{nhekmetric}) with $f$ replaced by $1 + r^2/r_0^2$.  This form of the metric is known as the NHEK geometry in global coordinates.  One notes that it has two boundaries, at $r = \pm \infty$.   Below, we will write {\it all} expressions in terms of the global coordinates $(t, r, \theta, \phi)$ used in (\ref{nhekglobal})  and we will set $r_0 = 1$.  It will sometimes be convenient to make the additional coordinate change $x \equiv \cos \theta$.  When making various expansions about the NHEK geometry, we will denote the metric (\ref{nhekglobal}) as the background metric $\bar g$.  One can also find other coordinates in which this same spacetime becomes a black hole of any finite temperature $T$ \cite{NoDynamics,Dias:2009ex}.

The throat geometry also inherits the isometries of AdS${}_2$.  These are well-known to form an $SL(2,\R)$ algebra and are given by
\begin{eqnarray}
\label{eat}
\eta_{-1} &=& \left( \frac{1}{2} - \frac{r \cos t}{2\sqrt{1+r^2}} \right) \,\partial_t - \frac{1}{2} \sin t \sqrt{1+r^2} \,\partial_r - \frac{\cos t}{2 \sqrt{1+r^2}}  \,\partial_\phi, \\
\eta_0 &=&  \frac{r \sin t}{\sqrt{1+r^2}} \, \partial_t - \cos t \sqrt{1+r^2}  \, \partial_r + \frac{\sin t}{\sqrt{1+r^2}} \, \partial_\phi\\
\eta_{1} &=&  \left( 1 + \frac{r \cos t}{\sqrt{1+r^2}} \right) \, \partial_t +  \sin t \sqrt{1+r^2} \, \partial_r + \frac{\cos t}{ \sqrt{1+r^2}} \, \partial_\phi,
\end{eqnarray}
in global coordinates.  The Lie brackets of these vector fields satisfy
\begin{equation}
\label{algebra}
[\eta_0, \eta_{\pm 1}] = \mp \eta_{\pm 1}, \ \ \ [\eta_{1}, \eta_{-1}] = \eta_0.
\end{equation}
There is also one additional Killing field, $\xi_0 = \partial_\phi$, which commutes  with all $\eta_i$.  The global time translation is $\partial_t = \frac{1}{2}\eta_1 + \eta_{-1}$.

\section{The covariant phase space}
\label{cps}

The basic idea of \cite{Guica:2008mu} was to consider a certain space of solutions asymptotic to the extreme Kerr throat (\ref{nhekmetric}) and to analyze the symmetries of the resulting phase space.  The space of solutions was determined by first specifying certain fall-off conditions at large $r$ of the form
\begin{equation}
\label{GHSS}
h_{ab}=
\left(\begin{array}{cccc}h_{tt}=\Op(r^{2}) & h_{tr}=\Op(r^{-2}) & h_{t\theta}=\Op(r^{-1})& h_{t\phi}=\Op(1) \\
& h_{rr}=\Op(r^{-3} )& h_{r\theta}=\Op(r^{-2}) & h_{r\phi}=\Op(r^{-1}) \\
&  & h_{\theta \theta}=\Op(r^{-1} )& h_{\theta\phi}=\Op(r^{-1}) \\
& &  & h_{\phi \phi}=\Op(1)
\end{array}\right) \,
\end{equation}
and then restricting to those solutions with vanishing energy, by which we mean the charge asymptotically associated with the vector field $\partial_t$.  The restriction on the energy was motivated in part by the desire to study ground states of the Kerr black hole, but also in part by the fact that (\ref{GHSS}) did not immediately guarantee that the energy was finite.  Restricting to zero-energy solutions was thus a way to avoid certain potential divergences.

This zero-energy condition was then analyzed further in \cite{NoDynamics,Dias:2009ex}, where it was argued to follow from the fall-off conditions (\ref{GHSS}).   We refer to (\ref{GHSS}) supplemented by this condition as GHSS boundary conditions. Ref. \cite{NoDynamics} argued that the other $SL(2, \mathbb{R})$ charges must vanish as well.   A similar statement was argued to hold for the $U(1)$ charge, with the exception that this charge can be carried by boundary gravitons.   Furthermore, \cite{NoDynamics,Dias:2009ex} showed that the only solutions to the linearized equations which satisfy at all times (\ref{GHSS}) and the constraint that these charges vanish are linearized diffeomorphisms\footnote{Interestingly, it is no problem to find initial data where the charges vanish.  The problem is that the charges are not conserved.  As a result, the Cauchy problem with GHSS boundary conditions does not appear to be well-defined. Classifying solutions is therefore not equivalent to classifying initial data.}.    In \cite{NoDynamics}, these conclusions were argued to hold under more general boundary conditions as well.  As a result, \cite{NoDynamics,Dias:2009ex} tentatively reached the conclusion that all solutions satisfying GHSS fall-off are diffeomorphic to (\ref{nhekmetric}).   We will use this conjecture to greatly simplify our analysis below.

We now turn to the construction of the covariant phase space.  Recall that the basic object in this procedure is the symplectic current $\omega^a$, which is closely connected with both the action and the equations of motion (see e.g. \cite{Iyer:1994ys,Iyer:1995kg,Wald:1999wa,Barnich:2001jy,Barnich:2007bf} for various discussions in the gravitational context).  The symplectic current is a vector field on spacetime, but it is also a two-form on the space of solutions, meaning that it depends on a point in the solution space (which we think of as a metric $g$) and two tangent vectors to this solution space at $g$.    Such tangent vectors $\delta_1 g$, $\delta_2 g$ are simply solutions to the  linearized equations of motion about $g$ which extend to one-parameter families of non-linear solutions; i.e., to curves in the specified phase space.   For each choice of linearized solutions $\delta_1 g$, $\delta_2 g$, the symplectic current is covariantly conserved: $\nabla_a \omega^a =0$, where $\nabla_a$ is the covariant derivative compatible with the metric $g$.

The integral of this current over a complete hypersurface $\Sigma$ defines the symplectic structure:
\begin{equation}
\label{OS}
\Omega_\Sigma[g; \delta_1g, \delta_2 g] = \int_\Sigma \sqrt{g_\Sigma} \, n_a \omega^a,
\end{equation}
where $g_\Sigma$ is the induced metric on $\Sigma$ and $n^a$ is the unit normal.  Covariant conservation of $\omega^a$ means that $\Omega_\Sigma-\Omega_{\Sigma'}$ can be related to the flux $\F$ of $\omega^a$ through any surface $\I$ for which $\partial \I = \partial \Sigma - \partial \Sigma'$; i.e., to the flux of $\omega^a$ through the boundary.

In order to define a good phase space,  (\ref{OS}) must be finite and conserved.  In particular,  the flux through the boundary must vanish.  These properties were not directly checked in \cite{Guica:2008mu}, though some related properties (finiteness and a form of integrability) were explored for certain charges constructed from $(\ref{OS})$ for the particular case of hypersurfaces $\Sigma$ on which $t = constant$.  We will discuss charges in section \ref{bst}, and concentrate for the moment on the symplectic structure itself.

As emphasized in \cite{Compere:2008us,Amsel:2009rr}, the correct choice of symplectic structure for a given problem can depend on the boundary conditions.  A useful way of organizing the discussion is to suppose that one has at hand the complete action for the system, including any boundary terms required to yield a well-defined variational principle.   In this case, \cite{Compere:2008us,Amsel:2009rr} show how to obtain a symplectic current $\omega^a$  for which the flux of this current through the boundary vanishes under the boundary conditions that make the action stationary.  Of course, changing the boundary conditions may require the addition of new boundary terms to the action.  The new action then defines a new symplectic current $\tilde \omega^a$ which differs from the original current $\omega^a$ by what is effectively also a ``boundary term.''  By this we mean that $ {\boldsymbol{\tilde \omega} } =\boldsymbol{\omega} - d {\bf B}$ for some $(d-2)$-form ${\bf B}$, where (in $d$ spacetime dimensions) $\boldsymbol{\omega}$ is the $(d-1)$-form dual to the symplectic current $\omega^a$.  Furthermore, just as the action may require certain boundary terms in order to render it finite, the correct boundary term ${\bf B}$ in the symplectic structure can also be critical to ensuring finiteness of $\boldsymbol{\tilde \omega}$.  Refs. \cite{Compere:2008us,Amsel:2009rr} explored  situations of this sort in anti-de Sitter space, where the well-known ``counter-terms'' in the action (see e.g. \cite{Henningson:1998gx,Balasubramanian:1999re,deHaro:2000xn,Bianchi:2001de, Bianchi:2001kw,Skenderis:2002wp}) lead to corresponding counter-terms in $\omega^a$.

In the NHEK context we do not yet have a complete useful variational principle from which to construct $\omega^a$; i.e., the analogue of holographic renormalization in AdS has not yet been performed.  We therefore cannot directly calculate a unique symplectic structure.  However, since we study vacuum general relativity, the bulk action is just the Einstein-Hilbert action.  As is well known \cite{Iyer:1994ys,Iyer:1995kg,Wald:1999wa}, this determines $\boldsymbol{\omega}$ up to the addition of a boundary term ($d{\bf B}$) of the sort discussed above.  The message we take from \cite{Compere:2008us} is that this boundary term may nevertheless be dictated by the choice of boundary conditions, and cannot be chosen arbitrarily.  In particular, the boundary term should be chosen so that the flux of $\omega$ through the boundary vanishes\footnote{In general, it would be better to say that one must choose the boundary conditions and the boundary term $d{\bf B}$ together so that this flux vanishes.  Here, however, we take the point of view that we wish to use precisely the GHSS fall-off conditions of \cite{Guica:2008mu} in order to have the same asymptotic symmetry group.}.

The particular symplectic structure used in \cite{Guica:2008mu} was that of \cite{Barnich:2007bf}:
\begin{equation}
\label{ss}
\omega^a[g; \delta_1g,\delta_2g]=-P^{abcdef}\left( \delta_2g_{cd} \nabla_b \delta_1 g_{ef}-(1\leftrightarrow 2)\right)\,,
\end{equation}
where
\begin{eqnarray}
\nonumber
P^{abcdef}&=&\frac{1}{32 \pi } \left(g^{ab} g^{e(c}g^{d)f}+g^{cd} g^{a(e}g^{f)b}+g^{ef} g^{a(c}g^{d)b} \right. \\
&& \left. \qquad\qquad-g^{ab} g^{cd}g^{ef}-g^{a(e} g^{f)(c}g^{d)b}-g^{a(c} g^{d)(e}g^{f)b}\right)\,.
\end{eqnarray}
A useful property is that, as reviewed in section \ref{bst}, for generally covariant theories  $\boldsymbol{\omega}$ must be an exact form whenever $\delta_2 g$ is the infinitesimal change of $g$ under a diffeomorphism (i.e., $\delta_2 g = \pounds_\xi g$) \cite{Iyer:1994ys,Iyer:1995kg,Wald:1999wa,Barnich:2001jy,Barnich:2007bf}.  In particular, for (\ref{ss}) we have \cite{Barnich:2001jy,Barnich:2007bf}
$\boldsymbol{\omega}[g;\delta_1 g, \pounds_\zeta g] = d {\bf k}_\zeta [\delta_1 g, g]$
where
\begin{eqnarray}
\label{formk}
{\bf k}_\zeta[\delta_1 g,g] &=& -\frac{1}{32 \pi}\epsilon_{abcd} \left[\zeta^d \nabla^c ( g^{ef} \delta_1 g_{ef}) -\zeta^d \nabla_f \delta_1 g^{c f} +\zeta_f \nabla^d \delta_1 g^{c f}+\frac{1}{2} ( g^{ef} \delta_1 g_{ef}) \nabla^d \zeta^c \right.\nonumber\\
&& \qquad\qquad\left.-\delta_1 g^{c f}\nabla_f \zeta^{d} +\frac{1}{2} \delta_1 g^{f c} (\nabla^d \zeta_f +\nabla_f \zeta^d)\right] dx^a \wedge dx^b \, .
\end{eqnarray}

The symplectic structure (\ref{ss}) corresponds to fixing the ambiguity $\boldsymbol{\omega} \rightarrow \boldsymbol{\omega} - d{\bf B}$ mentioned above in such a way that (\ref{ss}) can be constructed directly from the equations of motion without the need for a Lagrangian.  While this is an appealing property, there is no a priori guarantee that (\ref{ss}) is in fact appropriate for GHSS fall-off.

Indeed, it turns out that the corresponding $\Omega_\Sigma$ is not conserved.  Furthermore, $\Omega_\Sigma$ diverges unless $\Sigma$ asymptotically approaches a constant $t$ hypersurface.  To see this, we must calculate (\ref{ss}) at large $r$.  As in \cite{Guica:2008mu}, we shall confine ourselves to studying small perturbations about the NHEK solution $\bar g$.  This analysis is greatly simplified by assuming that, as argued in \cite{NoDynamics}, all solutions which obey the boundary conditions (\ref{GHSS}) are diffeomorphic to ({\ref{nhekmetric}).  It follows that the only tangent vectors to the space of solutions are linearized diffeomorphisms.  As shown in \cite{Guica:2008mu}, the linearized diffeomorphisms about $\bar g$ which preserve GHSS fall-off are of the form
\begin{equation}
\label{asSym}
\zeta = \xi_{\epsilon(\phi)} + c_1 \partial_t + c_2 \eta  + \chi_{sub}\,,
\end{equation}
at {\it each} boundary.  Here, we have defined
\begin{equation}
\label{vgen}
\xi_\epsilon = -r \epsilon'(\phi)\partial_r + \epsilon(\phi) \partial_\phi \quad \mathrm{for\,any\,function}\, \epsilon(\phi)\,,
\end{equation}
\begin{equation}
\eta = t \partial_t-r \partial_r \,,
\end{equation}
and the subleading terms take the form \cite{Guica:2008mu}
\begin{equation}
\label{sub}
\chi_{sub} = \frac{C^{\,t}(t,\theta, \phi)}{r^3} \, \partial_t+ C^{\,r}(t, \theta, \phi) \, \partial_r+\frac{C^{\,\theta}(t, \theta, \phi)}{r} \, \partial_\theta +\frac{C^{\,\phi}(t, \theta, \phi)}{r^2} \, \partial_\phi +\ldots
\end{equation}
 for some arbitrary functions $C^{\,t},C^{\,r}, C^{\,\theta}, C^{\,\phi}$.  We will comment further on the precise meaning of (\ref{sub}) in section \ref{bst}.
(The term involving $\eta$ seems to have been inadvertently omitted in \cite{Guica:2008mu}, but $\eta$ also respects their boundary conditions.)  As discussed in \cite{Guica:2008mu}, the non-trivial parts of (\ref{asSym}) are determined by the Virasoro terms (\ref{vgen}).

 We note in passing that other boundary conditions \cite{Matsuo:2009sj,Matsuo:2009pg}  have been suggested for Kerr/CFT which allow certain so-called ``right-moving'' Virasoro generators in addition to the ``left-movers'' above (\ref{vgen}).  However, the fact that these right-moving generators appear to contain arbitrary functions of time suggests that in a fully consistent framework they can define only gauge transformations.   In particular, on general grounds (see footnote \ref{tf} in section \ref{bst}), the corresponding charges should vanish identically in a theory with a conserved symplectic structure.  Consistency then requires any right-moving central charge to vanish.  As a result, we do not consider such boundary conditions further.

A general vector field preserving GHSS fall-off can be decomposed into one piece that vanishes as $r \to -\infty$ but approaches \eqref {asSym} as $r \to \infty$ and a second piece which vanishes as $r \to \infty$ but approaches \eqref{asSym} as $r \to -\infty$. Now, as noted above, one may write (\ref{ss}) in the form $\boldsymbol{\omega} = d {\bf k}$ where ${\bf k}$ is locally constructed (see eqn \ref{formk}) from $\zeta_2, \delta_1 g = \pounds_{\zeta_1} \bar g$, and $\bar g$.  Thus, the symplectic product $\Omega_\Sigma[\bar g; \pounds_{\zeta_1} \bar g, \pounds_{\zeta_2} \bar g]$ can be written as a local integral over $\partial \Sigma$.  It follows that $\Omega_\Sigma$ vanishes when $\zeta_1$ and $\zeta_2$ are supported at opposite boundaries.  Below, we choose counter-terms which preserve this very natural behavior.  It is therefore sufficient to confine our analysis to the boundary at $r\to \infty$, taking all  diffeomorphisms to be of the form \eqref{asSym} as $r \to \infty$ and to be trivial as $r \to -\infty$. These diffeomorphisms will define a non-trivial stress tensor at $r = + \infty$, but will not contribute to the stress tensor at $r = - \infty$.  Of course, one obtains analogous results if the boundaries are interchanged.

As noted in \cite{Guica:2008mu}, the vector fields (\ref{vgen}) are not smooth at the poles of the sphere ($\theta = 0, \pi$).  Nonetheless, there are sufficiently smooth vector fields $\zeta$ whose asymptotic behavior is given by (\ref{vgen}).  For definiteness,  we use the regulated Virasoro generators
\begin{equation}
\label{regvgen}
\tilde \xi_\epsilon =  R(r) \,\frac{r^2 \sin^2 \theta}{1+r^2 \sin^2 \theta} \left(\epsilon(\phi) \partial_\phi -r \epsilon'(\phi) \partial_r \right)
\end{equation}
in the calculations below.  Here $R(r)$ is any smooth function that tends to $1$ as $r\to \infty$ and to $0$ as $r\to -\infty$ with corrections of order ${\cal O}(r^{-1})$.

After some calculation, one finds that the symplectic flux due to a Virasoro diffeomorphism  through some region ${\cal R}$ of the $(t,\phi)$ cylinder at $r = \infty$ is
\begin{equation}
\label{div}
\F[\bar g; \pounds_{\tilde \xi_1} \bar g, \pounds_{\tilde \xi_2} \bar g]= \int_{{(t,\phi) \in {\cal R}}\atop  { r \to \infty}} dt d\phi \, \partial_\phi\left[ \Op(r \log r) +\ldots \right]\,,
\end{equation}
where we have already performed the $\theta$ integral before taking the $r\to \infty$ limit.
For regions $\I$ bounded by constant time surfaces, one can use the fact that the integrand is a total $\phi$ derivative to argue that it gives a vanishing contribution to the flux.   However, for generic ${\cal R}$ the flux diverges as $r\to \infty$.  Because $\F \sim \Omega_\Sigma-\Omega_{\Sigma'}$, it follows that  perturbations $\delta g = \pounds_{\tilde \xi} \bar g$ are not normalizeable with respect to the symplectic structure on generic surfaces.  Similar results hold for the unregulated generators (\ref{vgen}) and for regulators not of the form (\ref{regvgen}).

Thus, we see that (\ref{ss}) does not lead to a well-defined covariant phase space under GHSS fall-off.   One can of course choose to define the theory only on $t=constant$ slices.  However, were this the end of the story, it would suggest that any dual theory is non-local, as outgoing flux from one region of the boundary would in general be balanced only by ingoing flux in a completely different region\footnote{See \cite{Marolf:2007in} for a discussion of anti-de Sitter boundary conditions with similar properties.  In that case they correspond explicitly to non-local deformations of the dual field theory.}.  In particular, one would expect no covariantly conserved stress tensor, since the expected conserved quantities could not be defined on general surfaces.

On the other hand, as discussed above, it is natural to ask if we can define a renormalized symplectic structure $\boldsymbol{\omega}_{ren}$ by adding an exact form $d{\bf B}$ to $\boldsymbol\omega$ which removes both the outward flux and the divergence on generic hypersurfaces $\Sigma$.  Because any $\delta g$ in our phase space is generated by diffeomorphisms, it is clear on general principles that this is possible. Since $\boldsymbol{\omega}[g;  \delta g,\pounds_\zeta g] = d {\bf k}_\zeta[\delta g]$,  it would be legitimate in our case to take $\boldsymbol{\omega}_{ren} = \boldsymbol{\omega} - d {\bf B} = 0$ identically.  However, the resulting theory would be trivial.  In particular, one would lose all of the results from \cite{Guica:2008mu} concerning the asymptotic symmetry group and the central charge.

We would like to preserve such results.  To do so, it is useful to note that the actual computations in \cite{Guica:2008mu} were performed on hypersurfaces $\Sigma$ of constant $t$.  As a result, they involve only $\int_\Sigma \boldsymbol{\omega} = \int_{\partial \Sigma} {\bf k}$, and one sees that we will obtain the same result for any $\boldsymbol{\omega}_{ren} = d {\bf k}^{ren}$ so long as integrals of ${\bf k}$ and ${\bf k}^{ren}$ over $\partial \Sigma$ agree.  While $\partial \Sigma$ consists of two pieces, recall that we consider only vectors fields $\zeta$ which are trivial at $r \to - \infty$ so that this boundary will not contribute to the ${\bf k}$ integral and the same will be true for our ${\bf k}^{ren}$.

With this in mind, we note that the spacetime is foliated by (topological) 2-spheres $S^2_{r,t}$ on which $r,t$ are constant and define $\partial \Sigma$ as $\lim_{r \to \infty} S^2_{r,t}$.  We also note that the above-mentioned results from \cite{Guica:2008mu} depend only on the pull back $\boldsymbol{\lambda} = \lambda dx \wedge d \phi$ of  ${\bf k}$ to $S^2_{r,t}$.  Here we have used $x = \cos \theta$ and $\phi$ as coordinates on $S^2_{r,t}$.  We therefore choose ${\bf k}^{ren}$ to be essentially the lift of $\boldsymbol{\lambda}$ back to the NHEK geometry defined by the coordinate system $(t, r, x, \phi)$.  Furthermore, the subleading pieces $\chi_{sub}$ in (\ref{asSym}) should not contribute, as they have zero norm with respect to $\boldsymbol{\omega}$ on constant $t$ surfaces.  We therefore define
\begin{eqnarray}
\label{kren}
{\bf k}^{ren}_{\zeta_2} [\pounds_{\zeta_1} \bar g] &=&  \left( \lambda_{\tilde \xi_{\epsilon_2}}[ \pounds_{\tilde \xi_{\epsilon_1}} \bar g] -  \lambda_{\tilde \xi_{\epsilon_1}}[ \pounds_{\tilde \xi_{\epsilon_2}} \bar g] \right) \frac{ dx  \wedge d \phi}{2} \cr
 &=&  \left( (k_{\tilde \xi_{\epsilon_2} }[\pounds_{\tilde \xi_{\epsilon_1}} \bar g] )_{x \phi}  -  (k_{\tilde \xi_{\epsilon_1} }[\pounds_{\tilde \xi_{\epsilon_2}} \bar g] )_{x \phi} \right) \frac{ dx  \wedge d \phi}{2} ,
\end{eqnarray}
and
\begin{equation}
\label{B}
{\bf B}[\bar g;  \pounds_{\zeta_1} \bar g,\pounds_{\zeta_2} \bar g]  = \frac{1}{2}\left({\bf k}_{\zeta_2} [\pounds_{\zeta_1} \bar g]- {\bf k}_{\zeta_1} [\pounds_{\zeta_2} \bar g] \right) -
{\bf k}^{ren}_{\zeta_2} [\pounds_{\zeta_1} \bar g] \,.
\end{equation}
In (\ref{kren}), $\xi_{\epsilon_{1,2}}$ are chosen as in (\ref{regvgen}) and $\epsilon_{1,2}$ are defined by the decomposition (\ref{asSym}) of $\zeta_{1,2}$.  Note that ${\bf k}^{ren}_{\zeta_1} [\pounds_{\zeta_2} g]$ and thus $\boldsymbol{\omega}_{ren}[\pounds_{\zeta_1} g, \pounds_{\zeta_2} g]$ are manifestly antisymmetric in $\zeta_1,\zeta_2$.  For simplicity, we follow \cite{Guica:2008mu} in restricting attention to the symplectic structure and charges about the background $g= \bar g$.  However, because our entire phase space is generated from $\bar g$ by acting with diffeormophisms, we expect an appropriate notion of diffeomorphism covariance to define a unique extension of (\ref{kren}, \ref{B}) to general $g \neq \bar g$.

For a number of reasons, our construction of ${\bf B}$ is not covariant in the bulk. However, it may be thought of as becoming covariant ``at the boundary" where $r$ has a distinguished role as the coordinate orthogonal to the boundary, $t$ has a distinguished role through the fact that $\partial_t$ commutes with the Virasoro generators, and where $\chi_{sub}$ is in some sense trivial.  This is consistent with interpreting our ${\bf B}$ as arising from covariant counter-terms in an action:  such counter-terms would define ${\bf B}$ only at the boundary and leave the extension to the bulk arbitrary.

Now, by construction, the integrals of ${\bf k}$ and ${\bf k}^{ren}$ agree over each $S^2_{r,t}$.  However, because every object that defines $\lambda$ is independent of $t$, the pull-back of $\boldsymbol{\omega}_{ren} = d {\bf k}^{ren}$ to $r = constant$ surfaces vanishes, so that the flux ${\cal F}^{ren}$ of $\boldsymbol{\omega}_{ren}$ through the boundary vanishes identically.

\section{Charges and a stress tensor}
\label{bst}

Having found a counter-term that leads to a truly covariant phase space in section \ref{cps}, we are now in a position to construct a (canonical) boundary stress tensor. We begin by reviewing the connection between the symplectic structure and the various conserved charges that the boundary stress tensor should compute.

The fundamental property of the symplectic structure is that it is the ``inverse'' of the Poisson structure on the algebra of observables.  By this one means that the variation $\delta A$ of any observable $A$ under a general variation $\delta g$ of the solution (within the chosen phase space) must be of the form
\begin{equation}
\label{varyA}
\delta A = \Omega( \delta g, \delta_A g),
\end{equation}
where $\delta_A g$ is some tangent vector field on the space of solutions; i.e., it is a solution to the linearized equations of motion about $g$ for each $g$ that extends to one-parameter family of solutions.  Here we assume that $\Omega$ is conserved, so that no subscript $\Sigma$ is necessary.  The Poisson Bracket of two such observables is then
\begin{equation}
\label{PB}
[A,B] = \delta_A B;
\end{equation}
i.e., the derivative of $B$ along the specified vector field $\delta_A g$.

This fact has two important implications for our purposes below.  First, equations (\ref{varyA}) and (\ref{PB}) imply that generators of symmetries must satisfy certain differential equations on the space of solutions.  In particular, any charge $Q_{\xi}$ which generates motion along some vector field $\xi$ by definition satisfies $[Q_\xi,B] = \pounds_{\xi} B$, where $\pounds_\xi$ is the Lie derivative along $\xi$.  As a result, the first variation $\delta Q_\xi[g; \delta g]$ of $Q_\xi$ along some tangent vector $\delta g$ at $g$ must satisfy
\begin{equation}
\label{varyQ}
\delta Q_\xi [g; \delta g]= \Omega( \delta g,\pounds_\xi g).
\end{equation}

Second, note that any degeneracy in $\Omega$ leads to an ambiguity in $\delta_A g$.  Suppose for example that $\delta_0 g$ is a vector field for which $\Omega(\delta_0 g, \delta g) =0$ for all $\delta g$ tangent to the given phase space.  Then we have
$\delta A = \Omega(\delta_A g + \delta_0 g, \delta g)$ whenever $\delta_A g$ satisfies (\ref{varyA}).  As a result, (\ref{PB}) is well-defined only if $\delta_0 B =0$ for all observables $B$.  In other words, we must regard $\delta_0 g$ as a gauge transformation under which all observables must be invariant.  This notion of gauge transformation coincides with the expected results for familiar examples and is closely connected to ambiguities in solving the equations of motion.  In particular, conservation of $\Omega$ implies that any linearized solution $\delta_0 g$ which vanishes in an open neighborhood of some complete hypersurface $\Sigma$ is a degenerate direction of this sort\footnote{\label{tf}  It is for this reason that, as mentioned in section \ref{cps}, the right-moving charges of \cite{Matsuo:2009sj,Matsuo:2009pg} should vanish.}.

In practice, precisely which transformations are gauge symmetries can depend strongly on the choice of boundary conditions.  For example, in generally covariant theories $\boldsymbol{\omega}$ becomes and exact form when $\delta_2  g = \pounds_\xi g$ for any smooth vector field $\xi$.  As a result, $\Omega(\delta g,\pounds_\xi g)$ can be reduced to a boundary integral.  It follows immediately that any compactly-supported diffeomorphism is a gauge symmetry, though more analysis is required for diffeomorphisms which act non-trivially near the boundary.  When the result does not vanish, $\xi$ is said to generate a non-trivial asymptotic symmetry of the system.   Using the results of \cite{Barnich:2001jy,Barnich:2007bf}, the explicit form of such boundary integrals for vector fields $\zeta$ of the form (\ref{asSym}) was computed in  \cite{Guica:2008mu}.  The results depend only on $\epsilon (\phi)$, so that any $\zeta$ with $\epsilon =0$ is pure gauge, at least around the background $\bar g$.

Moreover, any solution $g$ continuously connected to $\bar g$ can be obtained by exponentiating the action of the vector fields (\ref{asSym}) on $\bar g$; i.e., by a finite diffeomorphism generated by a vector field $\zeta$ of the form (\ref{asSym}).  Since all $\xi_\epsilon$ commute with $\eta$ and $\partial_t$, these finite diffeomorphisms map $\eta, \partial_t$ to vector fields of the form $\eta+ \chi_{sub}$, $\partial_t + \chi_{sub}$, which define degenerate directions of $\boldsymbol{\omega}$ at $\bar g$.    But the symplectic structure is covariant, so that $\pounds_{\eta} g, \pounds_{\partial_t}g$ must be degenerate directions of $\boldsymbol{\omega}$ at any such $g$.  At least under the assumption that the solution space is connected, it follows that any vector field asymptotic to  $\eta$ or $\partial_t$ generates a gauge transformation.   In particular, we see that time-translations are pure gauge.  While we shall not make further use of this observation here, one wonders whether it will have further implications for the Kerr/CFT correspondence.

Combining (\ref{varyQ}) and the fact that $\boldsymbol{\omega}_{ren} = d {\bf k}^{ren}$ yields
\begin{equation}
\delta Q_\xi[g] = \int_{\partial \Sigma} {\bf  k}^{ren}_\xi[\delta g,g] \,,
\end{equation}
where
${\bf   k}^{ren}_\xi[\delta g,g] = {\bf  k}_\xi[\delta g,g] - {\bf B}[g,\delta g, \pounds_{\xi} g]$.
Finite charge differences can be obtained by integrating $\delta Q$ over a path in the space of metrics.  The finite charge will be independent of
this path if a certain integrability condition is satisfied:
\begin{equation}
\label{intGHSS}
\int_{\partial \Sigma} {\bf K}_\xi[g, h_1, h_2] = 0\,,
\end{equation}
where
\begin{equation}
{\bf K}_\xi[g, h_1, h_2] ={\bf  k}^{ren}_\xi[h_1,g+h_2] -{\bf  k}^{ren}_\xi[h_2 ,g+h_1] -{\bf  k}^{ren}_\xi[h_1-h_2 ,g] \,.
\end{equation}

This integrability condition can equivalently be expressed \cite{Wald:1999wa} in terms of the symplectic current as
\begin{equation}
\label{Wint}
\int_{\partial \Sigma} \zeta \cdot \boldsymbol{\omega}_{ren}[g; \delta_1 g, \delta_2 g] = 0 \,,
\end{equation}
where $(\zeta \cdot \boldsymbol{\omega})_{bc} = \zeta^a \omega_{abc}$.  Now, by construction $\boldsymbol{\omega}_{ren}$ does not depend on the subleading parts of $\zeta$.  This means that ${\bf k}^{ren}$ is independent of $t$ and   $\boldsymbol{\omega}_{ren} \sim \partial_r  \lambda dr \wedge dx \wedge d\phi$, where the $\sim$ denotes equality up to the anti-symmetrization in (\ref{kren}). Furthermore, since we consider normalizeable modes it is clear that $\partial_r \lambda \to 0$.   It turns out that $\lambda \sim const + {\cal O} (r^{-2})$  at each $(\theta, \phi)$.  However, the convergence is not uniform due to the formation of a tall, thin spike near each pole.  These spikes arise from the fact that as $r$ becomes large, the regulated vector fields (\ref{regvgen}) develop large derivatives near the poles.  However, a careful analysis shows that the integral over each spike remains small in the large $r$ limit.  We find that (\ref{Wint}) is satisfied (for $g =  \bar g$) so long as \begin{equation}
\zeta^r = r \alpha
\label{uniform}
\end{equation} where $\alpha$ is a smooth bounded function for large $r$.  This particular condition turns out to depend on the choice of $\theta$-dependence used in (\ref{regvgen}) to make the Virasoro generators well-behaved at the poles of the sphere.  Other regulators give similar results, but for slightly different classes of allowed $\chi_{sub}$.  In this sense our regularized symplectic structure depends somewhat on the choice of regulator, though all regulators appear to give the same central charge.   A related feature of the original symplectic structure $\boldsymbol{\omega}$ is discussed below.

We note that the condition (\ref{uniform}) is satisfied by the regulated Virasoro generators (\ref{regvgen}), and also by any repeated commutator of these fields.  However, it is not necessarily satisfied by all vector fields that are formally of the form (\ref{asSym}) if the terms represented by $(\ldots)$ are understood to be subleading only pointwise on each $S^2_{r,t}$ (and not in some more uniform sense).  It appears that only vector fields satisfying both (\ref{regvgen}) and (\ref{uniform}) should be regarded as pure gauge.

It turns out, however, that (\ref{Wint}) does not vanish for the original symplectic structure (\ref{ss}), even when $\Sigma$ is a constant time surface.   The issue is again a tall, thin spike near the pole similar to the one above, but which now turns out to be somewhat larger.  The difference is due to the $dx \wedge dr$ and $ dr \wedge d \phi$ terms in  ${ \bf k}$ which were removed in defining $\boldsymbol{\omega}_{ren}$.  The details of the spike depend on the regulator, but we find a non-zero result for a broad class of regulators that includes the one suggested in footnote 6 of \cite{Guica:2008mu}.  As a result, despite the results in their appendix B, we find that the charges are not integrable without the addition of our counter-terms.  We also find a non-zero result for the difference of two regulated Virasoro generators corresponding to the same Virasoro element, but having different regulators; e.g., for
\begin{equation}
\label{regvgendiff}
\tilde \xi_\epsilon = \,\left( \frac{r^2 \sin^2 \theta}{1+r^2 \sin^2 \theta} - \frac{r^2 \sin^4 \theta}{1+r^2 \sin^4 \theta} \right) \left(\epsilon(\phi) \partial_\phi -r \epsilon'(\phi) \partial_r \right) \,.
\end{equation}
As a result, one sees that different choices of regulator are not gauge equivalent with respect to the original symplectic structure $\boldsymbol{\omega}$.

Having shown that the charges defined by $\boldsymbol{\omega}_{ren}$ are integrable, we would now like to construct a conserved stress tensor that yields these charges.  The idea is to construct conserved currents in $t, \phi$ space
\begin{equation}
\label{jk}
(j_{\zeta_2})_i = \int_{-1}^1 (\tilde k^{ren}_{\zeta_2})_{i x}  dx |_{r \to \infty}\, \ \ \ {\rm for} \ \ i,j,\ldots = t, \phi,
\end{equation}
which are related to the associated  two-dimensional stress-energy tensor by
\begin{equation}
\label{jT}
(j_{\zeta_2})_i =  T_{ij}[\zeta_1] {\zeta_2}^j \, .
\end{equation}
Here $\tilde {\bf k}^{ren} = {\bf k}^{ren} + d{\bf A}$ for some smooth ${\bf A}$, so that
$\tilde {\bf k}^{ren}$ and  ${\bf k}^{ren}$ define identical charges.  The trick is to choose ${\bf A}$ so that (\ref{jT}) is satisfied; i.e., so that (\ref{jk}) can be expressed linearly in terms of ${\zeta_2}$ without taking derivatives.    An appropriate choice turns out to be
\begin{equation}
{\bf A}_{\zeta_2}[\bar g;\pounds_{\zeta_1} \bar g] =  -\frac{3R(r)}{64 \pi} \, (1-x^2) \, \left(\epsilon_1 \epsilon_2''-4\epsilon_1 \epsilon_2-2 \epsilon_1' \epsilon_2'+ 3 \epsilon_1'' \epsilon_2 \right)\,dx \, .
\end{equation}
The $x$-dependence is largely arbitrary, though it must have the correct integral over $\theta \in [0, \pi]$ and should vanish rapidly enough at the poles to make $d{\bf A}$ continuous.

From (\ref{jk}) one finds that only $j_\phi$ is non-zero, and the only non-zero component of the stress tensor  is
\begin{equation}
\label{stress}
T_{\phi \phi}[\zeta_1] = \frac{J}{2 \pi } \,\left(2 \epsilon_1'-\epsilon_1'''\right)\,,
\end{equation}
where we have restored factors previously set to one.   Note that, strictly speaking, we have computed the variation of the stress tensor for small fluctuations about the background $\bar g$ since ${\bf \tilde k}^{ren}$ computes the linearized charges about $g$.  To be precise, our $T_{ij}[\zeta]$ should be  related to the full boundary stress tensor $T_{ij}(g)$ for a solution $g$ by   $T_{\phi \phi}[\zeta_1] = \partial_\lambda [T_{\phi \phi}(\bar g + \lambda \pounds_{\zeta_1} \bar g)]|_{\lambda = 0}$.

\setcounter{equation}{0}

\section{Discussion}
\label{disc}

We showed above that the symplectic structure $\Omega_\Sigma$ used in \cite{Guica:2008mu} is finite and conserved only when evaluated on surfaces $\Sigma$ on which $t$ becomes asymptotically constant.  As a result, the phase space resulting from this symplectic structure is not covariant.   Were this the end of the story, it would suggest that any dual theory is defined only on $t=constant$ surfaces and, as a result, that it is highly non-local.  A related issue was a subtle failure of integrability for the charges defined in \cite{Guica:2008mu}, even on constant $t$ hypersurfaces.  As a result, the Virasoro generators were ill-defined at the non-linear level.

However, we found that both issues could be resolved by adding an appropriate local counter-term ${\bf B}$ to obtain a renormalized symplectic structure $\boldsymbol{\omega}_{ren} = \boldsymbol{\omega} - d{\bf B}$.  Since counter-terms in the symplectic structure result from counter-terms in the action needed to obtain a good variational principle  \cite{Compere:2008us}, our work may be considered a first step toward constructing a so-called holographically renormalized action for NHEK asymptotics.

Our counter-term is not covariant in the bulk as it required a choice of spheres $S^2_{t, r}$ and a volume form $dx \wedge d \phi$ on each sphere.  However, it becomes covariant on the boundary if this boundary is defined as the limit of constant $r$ surfaces.  The point is that the spheres $S^2_{t, r}$ and the associated volume form are uniquely-defined in the limit of large $r$ under the GHSS fall-off conditions.  This is consistent with interpreting our ${\bf B}$ as arising from counter-terms in a covariant action:  such counter-terms would define ${\bf B}$ only at the boundary and leave the extension to the bulk arbitrary. If such boundary terms can be found, they would provide a more elegant definition of $\boldsymbol{\omega}_{ren}$.

Our renormalized symplectic structure $\boldsymbol{\omega}_{ren}$ was explicitly constructed to give the same symplectic products as that of \cite{Guica:2008mu} when evaluated on constant $t$ hypersurfaces.  As a result, it defines both the asymptotic symmetry algebra and central charge reported in  \cite{Guica:2008mu}, in terms of which the Kerr black hole entropy can be expressed via Cardy's formula. However, the theory is now covariant, so that the same results are obtained on any hypersurface $\Sigma$.  In addition, the charges are now integrable and therefore exist at a non-linear level\footnote{\label{TSS} It may also be possible to construct integrable charges using the original symplectic structure by modifying the definition of the generators for $g \neq \bar g$.  See \cite{Banados:2005da} for an example of this sort.  However, such an approach cannot resolve the hypersurface dependence issue.}.

We then used $\boldsymbol{\omega}_{ren}$ to construct a differential boundary stress tensor $T_{ij}$, where $i,j$ run only over $t, \phi$.  Only the component $T_{\phi \phi}$ was non-zero, and the result was time-independent.  The vanishing of the time components $T_{tj}$ was associated with the fact that time-translations are pure gauge.

Note that the trace of $T_{ij}$ should vanish in any conformal field theory. In particular, since we computed the difference between the stress tensors of nearby states, any trace anomaly will cancel. It is interesting to ask what metrics $\gamma_{ij}$ in the dual theory could make $T_{ij}$ both traceless and conserved.  Due to the symmetries of the NHEK geometry, we assume $\gamma_{ij}$ is to be independent of $t, \phi$.  The necessary and sufficient condition is then simply $\gamma^{\phi \phi} =0$.    Since any such metric has $\gamma_{tt} =0$, we see  that the gauge direction $\partial_t$ should be a null direction in the dual theory.

It is useful to write the stress tensor in terms of null coordinates $x^\pm$ with respect to $\gamma_{ij}$.   One may set $x^- = \phi$ (so that $dx^-$ is null) and take $x^+$ to be an appropriate linear combination of $t$ and $\phi$ (so that $dx^+$ is null).   Note that since $(\partial_t)^i T_{ij} =0$, one finds $T_{+j} =0$; i.e., the only non-zero component is $T_{--}$.   Said another way, in the dual theory one can think of the Virasoro algebra as generating diffeomorphisms along the null direction $\frac{\partial}{\partial x^-}$, though the theory is chiral in the usual sense that generators along $\partial_t \propto \frac{\partial}{\partial x^+}$ vanish.

Since $x^- = \phi$ is periodic, the dual theory lives on a null cylinder and describes a discrete light cone quantum theory  as suggested by the arguments of \cite{Balasubramanian:2009bg}.  However, there are also important differences between the proposed Kerr/CFT correspondence of \cite{Guica:2008mu} and the picture described in \cite{Balasubramanian:2009bg} for extreme BTZ black holes.  Chief among these is that \cite{Guica:2008mu} found a central charge that depends on $J$ while the temperature is a constant $T =1 / 2\pi$.  In contrast, for the BTZ case the central charge is independent of $J$, so that all values of $J$ can be described by the same discrete light cone quantum theory.

Now the reader may recall that, assuming as above that the arguments of \cite{NoDynamics,Dias:2009ex} are correct and that the only solutions compatible with GHSS fall-off are diffeomorphic to the NHEK spacetime, the original NHEK symplectic structure $\boldsymbol{\omega}$ is an exact form $d{\bf k}$ on the phase space.  As a result, one can imagine adding boundary terms that would significantly change the results reported in \cite{Guica:2008mu}.  Indeed, at the level that we have studied the problem in this work, one may simply define ${\bf k}^{ren}_{\xi_1}[g, \pounds_{\xi_2} g] = f dx \wedge d\phi$  for {\em any} appropriately anti-symmetric smooth function $f[g,\xi_1,\xi_2]$, where $g$ is thought of as an element of the Virasoro group and $\xi_1,\xi_2$ are tangent vectors at $g$. In particular, one could choose ${\bf k}^{ren}$ to be independent of $J$, in closer analogy with \cite{Balasubramanian:2009bg}. However, we see no particularly natural constructions of this type and, in any case, without compensating changes in the Virasoro charge $L_0$, the predictions of Cardy's formula would no longer agree with the Bekenstein-Hawking entropy $S_{BH} \sim J$ of the Kerr black hole.

As is clear from the paragraph above, the current level of discussion leaves great ambiguity in the choice of symplectic structure, and thus in the physical results.  Our choice (\ref{kren}) has desirable features in that  it resolves the above-mentioned integrability and covariance issues while retaining the main results from \cite{Guica:2008mu}.  However, we gave no first-principles argument that would distinguish (\ref{kren}) from any other choice.  It would be much more satisfying if
(\ref{kren}) could be derived from a local covariant variational principle with boundary terms, which would then constitute a holographically renormalized action for Kerr/CFT.  One would hope that such a framework would both restrict the possible choices of symplectic structure and help to interpret the ambiguities that remain.

\section*{Acknowledgements}
The authors are grateful to Geoffrey Comp\`ere for many useful discussions during this work, as well as for sharing his Mathematica code.  We thank Gary Horowitz for motivating conversations and collaboration in the early stages of this work.  We also thank Tom Hartman, Andy Strominger, and Wei Song for a discussion of integrability and Tom Hartman and Andy Strominger for comments on an earlier draft.   This work was supported in part by the US National Science Foundation under Grant No.~PHY05-55669, and by funds from the University of California.


\begin{thebibliography}{99}

\bibitem{Maldacena:1997re}
 J.~M.~Maldacena,
 ``The large N limit of superconformal field theories and supergravity,''
 Adv.\ Theor.\ Math.\ Phys.\  {\bf 2}, 231 (1998)
 [Int.\ J.\ Theor.\ Phys.\  {\bf 38}, 1113 (1999)]
 [arXiv:hep-th/9711200].


\bibitem{Zaslavsky:1997uu}
  O.~B.~Zaslavsky,
  ``Horizon/Matter Systems Near the Extreme State,''
  Class.\ Quant.\ Grav.\  {\bf 15}, 3251 (1998)
  [arXiv:gr-qc/9712007].


\bibitem{Bardeen:1999px}
 J.~M.~Bardeen and G.~T.~Horowitz,
 ``The extreme Kerr throat geometry: A vacuum analog of AdS(2) x S(2),''
 Phys.\ Rev.\  D {\bf 60}, 104030 (1999)
 [arXiv:hep-th/9905099].


\bibitem{Guica:2008mu}
 M.~Guica, T.~Hartman, W.~Song and A.~Strominger,
 ``The Kerr/CFT Correspondence,''
 arXiv:0809.4266 [hep-th].




\bibitem{Azeyanagi:2009wf}
T.~Azeyanagi, G.~Comp\`ere, N.~Ogawa, Y.~Tachikawa and S.~Terashima,
  ``Higher-Derivative Corrections to the Asymptotic Virasoro Symmetry of 4d
  Extremal Black Holes,''
  arXiv:0903.4176 [hep-th].


\bibitem{Balasubramanian:2009bg}
  V.~Balasubramanian, J.~de Boer, M.~M.~Sheikh-Jabbari and J.~Simon,
  ``What is a chiral 2d CFT? And what does it have to do with extremal black
  holes?,''
  arXiv:0906.3272 [hep-th].


\bibitem{Bredberg:2009pv}
  I.~Bredberg, T.~Hartman, W.~Song and A.~Strominger,
  ``Black Hole Superradiance From Kerr/CFT,''
  arXiv:0907.3477 [hep-th].


\bibitem{Compere:2008us}
 G.~Comp\`ere and D.~Marolf,
 ``Setting the boundary free in AdS/CFT,''
 Class.\ Quant.\ Grav.\  {\bf 25}, 195014 (2008)
 [arXiv:0805.1902 [hep-th]].

\bibitem{Amsel:2009rr}
  A.~J.~Amsel and G.~Comp\`ere,
  ``Supergravity at the boundary of AdS supergravity,''
  Phys.\ Rev.\ D {\bf 79}, 085006 (2009) [arXiv:0901.3609 [hep-th]].


\bibitem{Henningson:1998gx}
  M.~Henningson and K.~Skenderis,
  ``The holographic Weyl anomaly,''
  JHEP {\bf 9807}, 023 (1998)
  [arXiv:hep-th/9806087].




\bibitem{Balasubramanian:1999re}
V.~Balasubramanian and P.~Kraus, ``A stress tensor for anti-de {S}itter
  gravity,'' Commun. Math. Phys. {\bf 208} (1999) 413--428,
[arXiv:hep-th/9902121].

  \bibitem{deHaro:2000xn}
S.~de~Haro, S.~N. Solodukhin, and K.~Skenderis,
``Holographic reconstruction of spacetime and renormalization in the AdS/CFT correspondence,''
  Commun. Math. Phys. {\bf 217} (2001) 595--622, [arXiv:hep-th/0002230].

\bibitem{Bianchi:2001de}
M.~Bianchi, D.~Z. Freedman, and K.~Skenderis,
``How to go with an RG flow,''
  JHEP {\bf 08} (2001) 041, [arXiv:hep-th/0105276].

\bibitem{Bianchi:2001kw}
M.~Bianchi, D.~Z. Freedman, and K.~Skenderis,
``Holographic Renormalization,'' Nucl. Phys. {\bf B631} (2002) 159--194,
[arXiv:hep-th/0112119].

 \bibitem{Skenderis:2002wp}
  K.~Skenderis,
  ``Lecture notes on holographic renormalization,''
  Class.\ Quant.\ Grav.\  {\bf 19}, 5849 (2002)
  [arXiv:hep-th/0209067].



\bibitem{LL}
L. D. Landau and E. M. Lifshitz
{\it The Classical Theory of Fields,}
(Butterworth-Heinemann, London, 1980).



\bibitem{NoDynamics}
  A.~J.~Amsel, G.~T.~Horowitz, D.~Marolf and M.~M.~Roberts,
  ``No Dynamics in the Extremal Kerr Throat,''
  arXiv:0906.2376 [hep-th].

\bibitem{Dias:2009ex}
  O.~J.~C.~Dias, H.~S.~Reall and J.~E.~Santos,
  ``Kerr-CFT and gravitational perturbations,''
  arXiv:0906.2380 [hep-th].


\bibitem{Iyer:1994ys}
  V.~Iyer and R.~M.~Wald,
  ``Some properties of Noether charge and a proposal for dynamical black hole
  entropy,''
  Phys.\ Rev.\  D {\bf 50}, 846 (1994)
  [arXiv:gr-qc/9403028].


\bibitem{Iyer:1995kg}
  V.~Iyer and R.~M.~Wald,
  ``A Comparison of Noether charge and Euclidean methods for computing the
  entropy of stationary black holes,''
  Phys.\ Rev.\  D {\bf 52}, 4430 (1995)
  [arXiv:gr-qc/9503052].



\bibitem{Wald:1999wa}
 R.~M.~Wald and A.~Zoupas,
 ``A General Definition of ``Conserved Quantities'' in General Relativity and
 Other Theories of Gravity,''
 Phys.\ Rev.\  D {\bf 61}, 084027 (2000)
 [arXiv:gr-qc/9911095].


\bibitem{Barnich:2001jy}
 G.~Barnich and F.~Brandt,
 ``Covariant theory of asymptotic symmetries, conservation laws and  central
 charges,''
 Nucl.\ Phys.\  B {\bf 633}, 3 (2002)
 [arXiv:hep-th/0111246].


\bibitem{Barnich:2007bf}
 G.~Barnich and G.~Comp\`ere,
 ``Surface charge algebra in gauge theories and thermodynamic integrability,''
 J.\ Math.\ Phys.\  {\bf 49}, 042901 (2008)
 [arXiv:0708.2378 [gr-qc]].

\bibitem{Matsuo:2009sj}
  Y.~Matsuo, T.~Tsukioka and C.~M.~Yoo,
  ``Another Realization of Kerr/CFT Correspondence,''
  arXiv:0907.0303 [hep-th].

\bibitem{Matsuo:2009pg}
 Y.~Matsuo, T.~Tsukioka and C.~M.~Yoo,
  ``Yet Another Realization of Kerr/CFT Correspondence,''
  arXiv:0907.4272 [hep-th].



\bibitem{Marolf:2007in}
  D.~Marolf and S.~F.~Ross,
  ``Reversing Renormalization-Group Flows with AdS/CFT,''
  JHEP {\bf 0805}, 055 (2008)
  [arXiv:0705.4642 [hep-th]].



\bibitem{Banados:2005da}
  M.~Banados, G.~Barnich, G.~Comp\`ere and A.~Gomberoff,
  ``Three dimensional origin of Goedel spacetimes and black holes,''
  Phys.\ Rev.\  D {\bf 73}, 044006 (2006)
  [arXiv:hep-th/0512105].


\end{thebibliography}
\end{document}